\documentclass{aa} 
\usepackage{graphicx}
\usepackage{epstopdf}
\usepackage{natbib}
\usepackage{txfonts}
\usepackage{color}
\usepackage[switch, modulo]{lineno}

\begin{document}

\title{High-precision broad-band linear polarimetry of early-type binaries}
                                
\subtitle{I. Discovery of variable, phase-locked polarization in HD 48099}

\author{A. Berdyugin
         \inst{1}     
         \and  V. Piirola \inst{2,4}
         \and S. Sadegi \inst{1} 
         \and  S. Tsygankov \inst{1}   
         \and T. Sakanoi
         \inst{3}
         \and M. Kagitani
          \inst{3} 
         \and  M. Yoneda
          \inst{4}
         \and  S. Okano
          \inst{3}
        \and    J. Poutanen
          \inst{1}        
          }

   \institute{Tuorla Observatory, Department of Physics and Astronomy,  University of Turku, 
              V\"ais\"al\"antie 20, 21500 Piikki\"o, Finland
       \and
             FINCA, University of Turku,  V\"ais\"al\"antie 20, 21500 Piikki\"o, Finland
       \and
             Graduate School of Science, Tohoku University,
              Aoba-ku, Sendai 980-8578 Japan            
       \and 
             Kiepenheuer-Institut f\"{u}r Sonnenphysik, D-79104 Freiburg, Germany
       \      
          }
\titlerunning{High-precision linear polarimetry of early-type binaries}
\authorrunning{A. Berdyugin et al. }
\date{Received; accepted}


\abstract
{}
{We investigate the structure of the O-type binary system HD 48099 by measuring linear polarization that arises due to light scattering process. High-precison polarimetry provides independent estimates of the orbital parameters and gives important information on the properties of the system.}
{Linear polarization measurements of HD 48099 in the B, V and R passbands with the high-precision Dipol-2 polarimeter have been carried out. The data have been obtained with the 60 cm KVA (Observatory Roque de los Muchachos, La Palma, Spain) and T60 (Haleakala, Hawaii, USA) remotely controlled telescopes during 31 observing nights.
Polarimetry in the optical wavelengths has been complemented by observations in the X-rays with the  {\it SWIFT} space observatory.}   
{Optical polarimetry revealed small intrinsic polarization in HD 48099 with $\sim 0.1\%$ peak to peak variation over the orbital period of 3.08~d. The variability pattern is typical for binary systems, showing strong second harmonic of the orbital period. We apply our model code for the electron scattering in the circumstellar matter to put constraints on the system geometry. A good model fit is obtained for scattering of light on a cloud produced by the colliding stellar winds. The geometry of the cloud, with a broad distribution of scattering particles away from the orbital plane, helps in constraining the (low)  orbital inclination. We derive from the polarization data the inclination $i = 17\degr \pm 2\degr$ and the longitude of the ascending node  $\Omega = 82\degr \pm 1\degr$ of the binary orbit. The available X-ray data provide additional evidence for the existence of the colliding stellar winds in the system. Another possible source of the polarized light could be scattering from the stellar photospheres. The models with circumstellar envelopes, or matter confined to the orbital plane, do not provide good constraints on the low inclination, better than $i \le 27\degr$, as is already suggested by the absence of eclipses.}
{}

   
\maketitle
   
%

\section{Introduction}

The scattering of light in close binary system may result in variable linear polarization which is synchronous with the orbital motion. Studying this polarization allows us to draw conclusions about the properties of the binary components and the structure of the circum\-binary environment. Polarimetry can also provide estimates of the orbit inclination $i$ and the longitude of the ascending node  $\Omega$ \citep[see][]{bme78,drissen86}. Independent estimates of the inclination are especially important for non-eclipsing systems for which spectroscopy and photometry alone do not allow us to obtain it directly from the observational data. The unknown inclination introduces uncertainty in the component masses and has an adverse effect on estimates of other stellar parameters. Therefore, in the case of binary stars, polarimetry may serve as a useful additional tool with good diagnostic capability.            
                                
In early-type binaries, variable phase-locked polarization usually arises from the electron scattering from gaseous envelopes, disks, streams or a dense stellar wind. Scattering (reflection) of the incident radiation from the close companions could also be responsible for the observed polarization variability \citep{berd99}. The magnitude of polarization depends, in general, on the number of free electrons in the system and the geometrical properties of the light scattering region. In WR+O systems with dense stellar wind, like V444 Cyg, the amplitude of variability in broad-band polarization can be as large as $\ge 0.3\%$ \citep[see][]{st-louis93}. In detached O + O systems with tenuous wind ($\dot M \le 10^{-7}M_{\odot}{\rm yr}^{-1}$), the expected amplitude could be much lower, at the level of  $\le 0.1\%$ ($\le 10^{-3}$). Most of these binaries do not show spectroscopic traces of H$_{\alpha}$ emission which is considered as a signature of the envelope, accretion stream or colliding stellar winds (see, for example, \citealt{thaller97}). Therefore, high precision measurements are needed to detect and study polarization variability in such systems.   
   
Recently, we have developed and built the Dipol-2 polarimeter that is capable of measuring linear polarization in the B, V and R-bands with the precision of $10^{-5}$ \citep{piirola14}. We use this instrument in research projects where high-precision polarimetry is required. In 2012, we began an observational campaign on the polarimetry  of early-type binaries, with the remotely controlled 60 cm KVA telescope (La Palma, Spain). In this paper we present our first results obtained for the detached non-eclipsing system HD 48099.     
   
\subsection{HD 48099}   
           \label{sec:source}
           
HD 48099 is a double-line spectroscopic binary located near the open cluster NGC 2244. The orbital period is  $P_{\rm orb} = 3.07806$~d and the binary consists of an 05.5V ((f)) primary and an O9V secondary with $M_{1}\sin^3i = 0.70M_\odot$ and $M_{2}\sin^3i = 0.39M_\odot$ (see  \citet{mahy10} for the recent extensive spectroscopic study of the system). The measured semi-amplitudes of the radial velocities variations of $K_1=54.4$~km~s$^{-1}$  and $K_2=96.2$~km~s$^{-1}$ give the following semimajor axes of the stellar orbits around centre of mass  $a_{1} \sin i = 3.31 R_{\odot}$ and $a_{2} \sin i = 5.85 R_{\odot}$.
The small minimum masses and low semi-amplitudes of radial velocity variations suggest a small orbit inclination and this accounts for the absence of eclipses in HD 48099. Modelling stellar spectra yielded the radii of the components:  
$R_{1} = 11.6 R_{\odot}$ and $R_{2} = 6.5 R_{\odot}$. 
The estimated stellar sizes imply that the eclipses should already occur at $i \ge 27\degr$. 

One of the most striking features of this system is a surprisingly small luminosity difference between the primary and the secondary in the UV wavelengths measured by \citet{stickland96}, who found a value of $l_{1}/l_{2} = 1.8$. This is in sharp contrast with the luminosity ratio in the optical $l_{1}/l_{2}= 3.96$ derived by \citet{mahy10}. The determined spectral types would imply even larger luminosity ratio in the UV than in the optical. The primary star in HD 48099 appears to be a fast rotator with strongly clumped stellar wind and enhanced nitrogen abundance \citep{mahy10}. 

Although HD 48099 has never been systematically studied for brightness variations, quite extensive photometry data have been obtained for this star by the Hipparcos satellite \citep{1997ESASP1200.....E}. We have retrieved these data via SIMBAD Astronomical Database and in Figure 1 we plotted the stellar brightness measured in Hipparcos `H' magnitude versus orbital phase using the ephemeris from  \citet{mahy10},  with the time  of the primary conjunction $T_{0}$ = HJD 2452649.661.
As we see in the plot, the available light curve, apart from being noisy, does not show any signatures of even partial eclipses. However, there is a hint at weak maxima of brightness near the elongations (phases 0.25 and 0.75), which is typical for the ellipsoidal variables; this may suggest that one or both components in HD 48099 are tidally distorted.      

\begin{figure} 
\centering
\includegraphics[width=9cm]{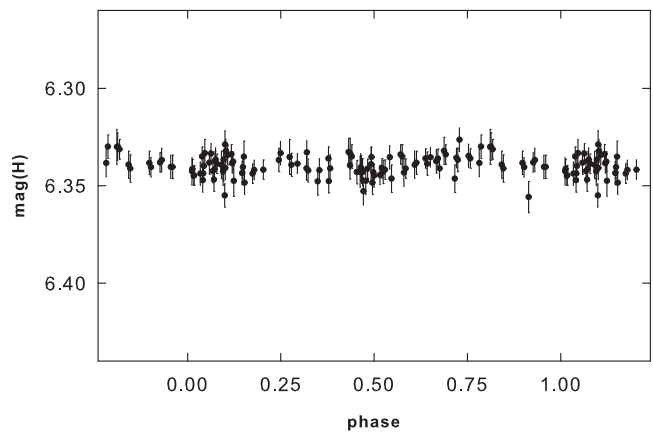}
\caption{Hipparcos stellar magnitude `H' of HD 48099 plotted against the phase of the orbital period.}
\label{Fig01}%
\end{figure}

\section{Optical polarimetry} 
   
\subsection{Polarimetric observations}
   
Dipol-2 polarimeter is equipped with three CCD cameras and two dichroic mirrors allowing polarimetric measurements in the B, V and R bands simultaneously. The superachromatic half-wave plate rotated by a stepper motor is used as  polarization modulator and a plane-parallel calcite plate provides separation of two orthogonally polarized stellar images in the focal plane. These images are recorded by the CCD cameras and then extracted for polarization analysis. Three copies of Dipol-2 have been built so far. Detailed description of the instrument is given by  \citet{piirola14}.     

   \begin{table} 
      \caption{Instrumental Polarization: KVA} 
   \label{InstPolKVA} 
   \centering 
   \begin{tabular}{c c c} 
   \hline\hline 
    Band & $q_{\rm inst}(\%)$ & $u_{\rm inst}(\%)$ \\ 
   \hline 
   B & $0.0069 \pm 0.0004$ & $0.0069 \pm 0.0003$ \\ 
   V & $0.0043 \pm 0.0003$ & $0.0059 \pm 0.0003$  \\
   R & $0.0033 \pm 0.0003$ & $0.0044 \pm 0.0003$ \\
   \hline 
   \end{tabular}
   \end{table}  
               
   \begin{table} 
   \caption{Instrumental Polarization: T60} 
   \label{InstPolT60} 
   \centering 
   \begin{tabular}{c c c} 
   \hline\hline 
   Band & $q_{\rm inst}(\%)$ & $u_{\rm inst}(\%)$ \\ 
   \hline 
   B & $-0.0007 \pm 0.0002$ & $-0.0007 \pm 0.0002$ \\ 
   V & $0.0002 \pm 0.0002$ & $-0.0005 \pm 0.0002$  \\
   R & $0.0010 \pm 0.0002$ & $-0.0006 \pm 0.0002$ \\
   \hline 
   \end{tabular}
   \end{table}                 
                
\begin{figure*} 
 \centering
 \includegraphics[width=18cm]{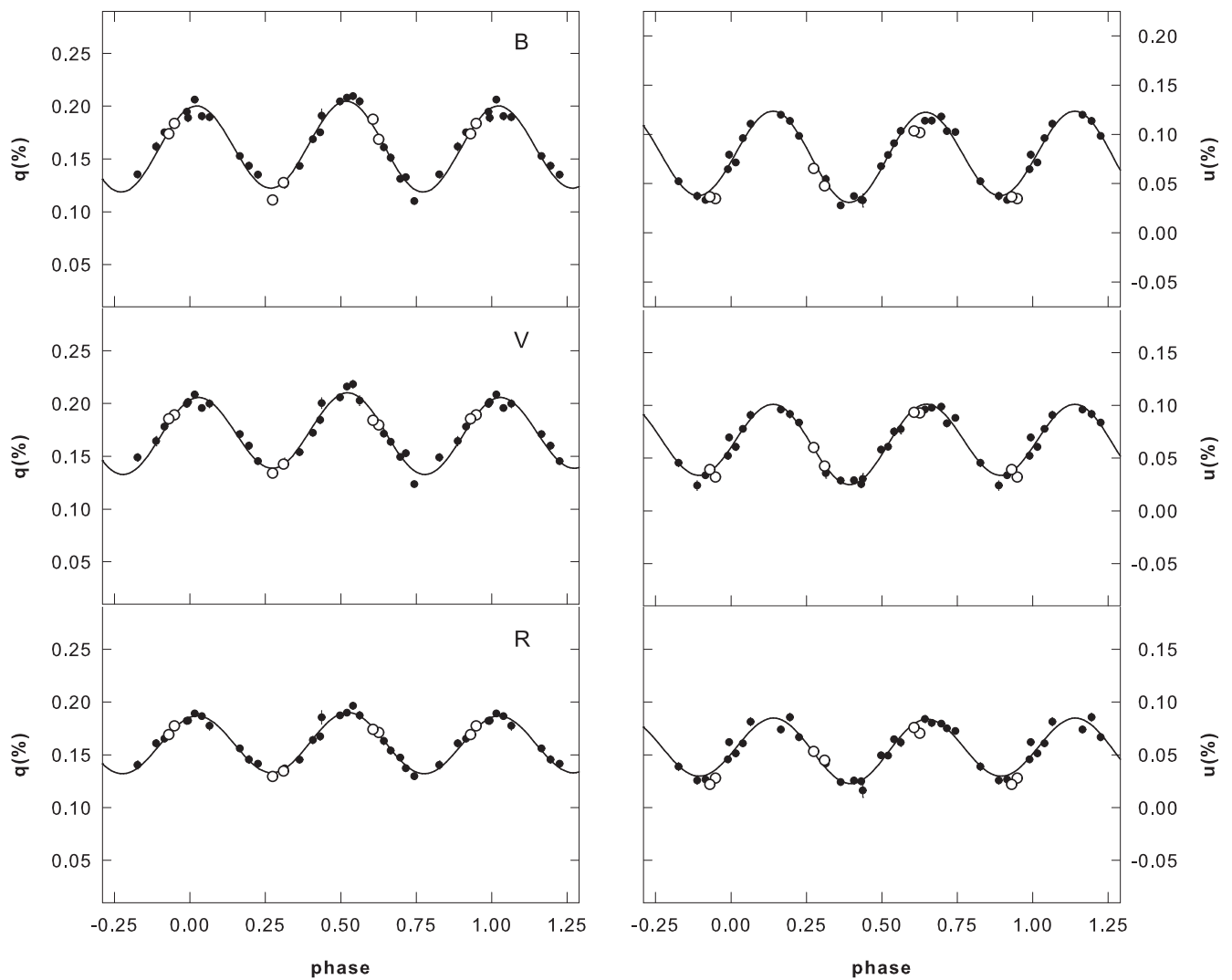}
 \caption{Stokes parameters $q$ and $u$ of HD 48099 plotted against the orbital period phase. The data obtained with the KVA and T60 telescopes are shown by filled and hollow circles respectively. Phase `0.0' corresponds to primary conjunction. The Fourier fit with two harmonics 
 is indicated by the solid line. Error bars ($\pm\sigma$) are shown, or are less than the size of the symbol.}
 \label{Fig02}%
\end{figure*}               
               
Our linear polarimetry of HD 48099 was carried out in 2013--2015. Most of the data (25 measurements) were obtained on the 60 cm KVA telescope at Observatorio del Roque de los Muchachos, La Palma in 2013--2014. In January 2015, additional data (six measurements) were acquired at the T60 telescope at Haleakala Observatory, Hawaii with the second copy of the polarimeter.   

Removal of the instrumental polarization $P_{\rm inst}$ has been done by observations of zero-polarized standard stars. More than 20 stars have been measured on both telescopes and the average values of instrumental polarization in the B, V and R-bands have been computed. Zero-point of the polarization angle has been determined by the observations of the highly polarized standards HD 25443, HD 161056 and HD 204827.
  
Tables~\ref{InstPolKVA} and \ref{InstPolT60} show the values of Stokes parameters $q_{\rm inst}$ and $u_{\rm inst}$ of the instrumental polarization. We see that for both telescopes, $P_{\rm inst} < 0.01\%$. The high precision that has been achieved in the determination of instrumental polarization (at the level of $10^{-6}$) allows us to calibrate our measurements thoroughly and guarantees consistency between the data obtained on two different telescopes.                      
   
The total duration of polarization measuring cycle for HD 48099 was about of 1--1.5 hours per night. During this time, from 48 to 64 single measurements of Stokes parameters $q$ and $u$ were obtained and the average nightly values of polarization were computed. Typical errors are in the range of $0.003\% - 0.005\%$, being dependent mostly on photon noise in the given passband and sky conditions. 
Polarization data obtained for HD 48099 are presented in Table 5\footnote{Table 5 with the polarimetry data is only available in electronic form at the CDS via anonymous ftp to cdsarc.u-strasbg.fr (130.79.128.5) or via http://cdsweb.u-strasbg.fr/cgi-bin/qcat?/A+A/}. 
 
\begin{table*}
\begin{center}
\caption[Swift/XRT observations of the source HD48099]{Swift/XRT observations of the source HD48099} \label{tablog48} 
\begin{tabular}{cccccc}
\hline
\hline
Obs Id &  Date & Date     & Orbital phase & Exposure & XRT count rate$^{a}$ \\
       &       & HJD       &    & s          & cts s$^{-1}$         \\
\hline
00033711001 & 2015-03-31T01:52:33 & 57112.0834 & 0.914 & 1031.4 & $(4.4\pm2.2)\times10^{-3}$  \\
00033711005 & 2015-03-31T17:47:19 & 57112.7490 & 0.130 & 1440.9 & $(8.8\pm2.6)\times10^{-3}$  \\
00033711003 & 2015-04-01T00:20:15 & 57113.0200 & 0.218 & 1101.3 & $(16.7\pm4.0)\times10^{-3}$  \\
00033711004 & 2015-04-01T22:49:56 & 57113.9568 & 0.523 & 1061.3 & $(11.7\pm3.4)\times10^{-3}$  \\
\hline
\end{tabular}
\end{center}
\begin{center}{\small \textbf{Notes:} $^{a}$  Net count rate in 0.3--10 keV energy range. }
\end{center}
\end{table*}

 \subsection{Polarization variability}

Our measurements clearly revealed low-amplitude ($\Delta P \le 0.1\%$) phase-locked variability of polarization in HD 48099. The variations of Stokes parameters $q$ and $u$ with the orbital phase are shown in Figure~\ref{Fig02}, and are available online at the CDS. This figure also shows the Fourier fit to the data, including first and second harmonics of the orbital period. The second harmonic variations clearly dominate. According to the analytic scattering models of \citet{bme78} (hereafter the BME model), this is typical for a binary system where the light scattering material is symmetrically distributed about the orbital plane. The average values of polarization are small in all passbands: $P \le 0.25\%$. This shows that the interstellar polarization in the direction of HD 48099 is small despite of substantial reddening: E(B-V) = 0.223 \citep{mahy10}.

The amplitude of the variations decreases from $\simeq0.10\%$ in the B to $\simeq0.07\%$ in the R passband. Although Thomson scattering is a grey process, this decrease of polarization can be explained as being due to the dilution effect of unpolarized free-free emission from the surrounding circumstellar matter, increasing towards longer wavelengths. As seen in Figure~\ref{Fig02}, the amplitudes of the variations in Stokes $q$ and $u$ are nearly equal. Again, if we interpret this in the context of analytical models for symmetric circumstellar envelopes, this is an indication of low inclination of the binary orbit  \citep[see Fig.~7 in ][]{bme78}.

\section{X-ray observations}
\label{sec:xrays}

To test the possibility that the system may contain colliding winds, we arranged observations of HD 48099 with the  {\it Swift} X-ray space observatory. {\it XRT} telescope onboard the {\it Swift} observatory  \citep{gehr2004} provides the possibility of observing sources of X-ray emission between 0.3 and 10 keV. We used the data from four TOO observations of HD 48099 covering one orbital cycle between March 31 and April 1, 2015. The observation log is presented in Table~\ref{tablog48}. All measurements were done in Photon Counting (PC) mode and the data were reduced using standard pipeline following \citet{evans2009}. Further data analysis was done using tools and packages available in {\sc ftools/heasoft} version 6.15.1.

\begin{figure}
 \includegraphics[width=8cm]{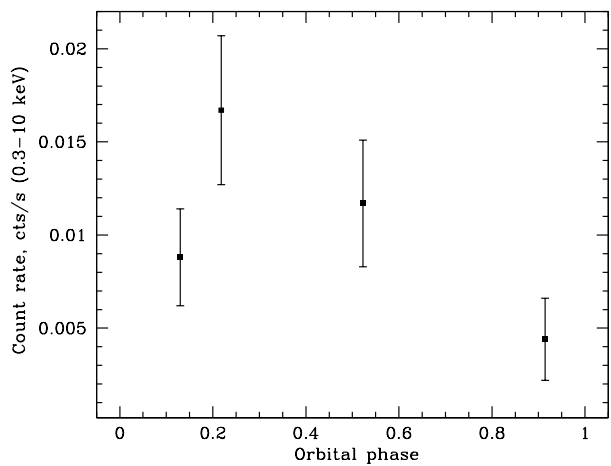}
 \caption{Count rate of HD48099 in 0.3--10 keV energy range obtained by the {\it Swift}/XRT telescope as a function of the orbital phase.}
 \label{lc48}
\end{figure}      

\begin{figure}
\includegraphics[width=8cm]{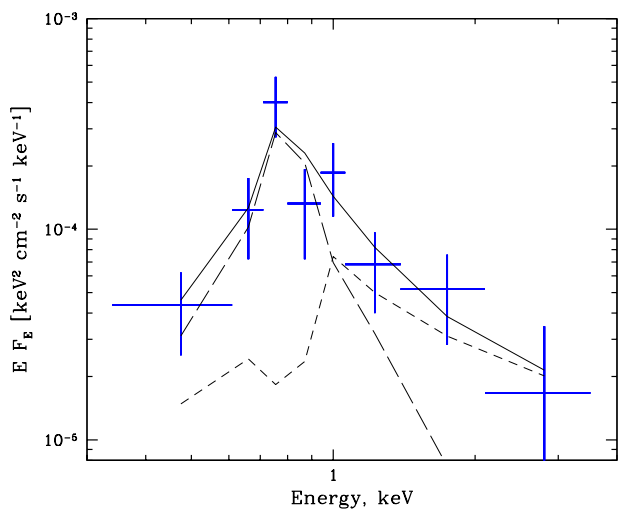}
 \caption{
 {\it Swift}/XRT spectrum of HD 48099 averaged over all available observations. The solid line represents the best fit model consisting of two {\sc APEC} components with temperatures $\sim0.4$ and $\sim1.7$ keV shown separately with long-dashed and short-dashed lines, respectively.}
 \label{x-rays_sp}
\end{figure}

\begin{figure} 
\centering
 \includegraphics[width=9cm]{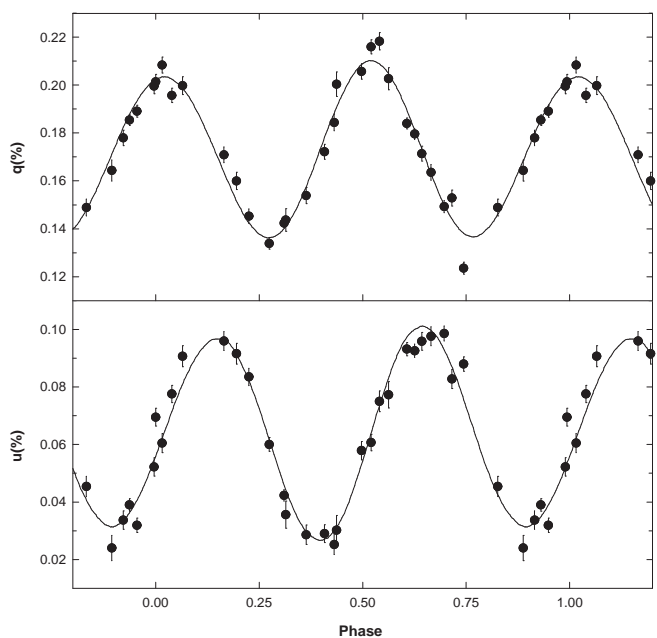}
 \caption{Best model fit to the polarization data in the V-band for the reflecting shells scenario. 
 The ellipsoidal shell around the primary has the equatorial radius  $r_{e} = 1.1 R_{A}$ and polar radius
 $r_{p} = 1.05 R_{A}$, where $R_{A}$ is the radius of the primary star. Spherical illuminating stars with the linear limb darkening coefficient 0.2 have been assumed.}
 \label{shells_fit}%
\end{figure}

\begin{figure*} 
\centering
\includegraphics[width=18cm]{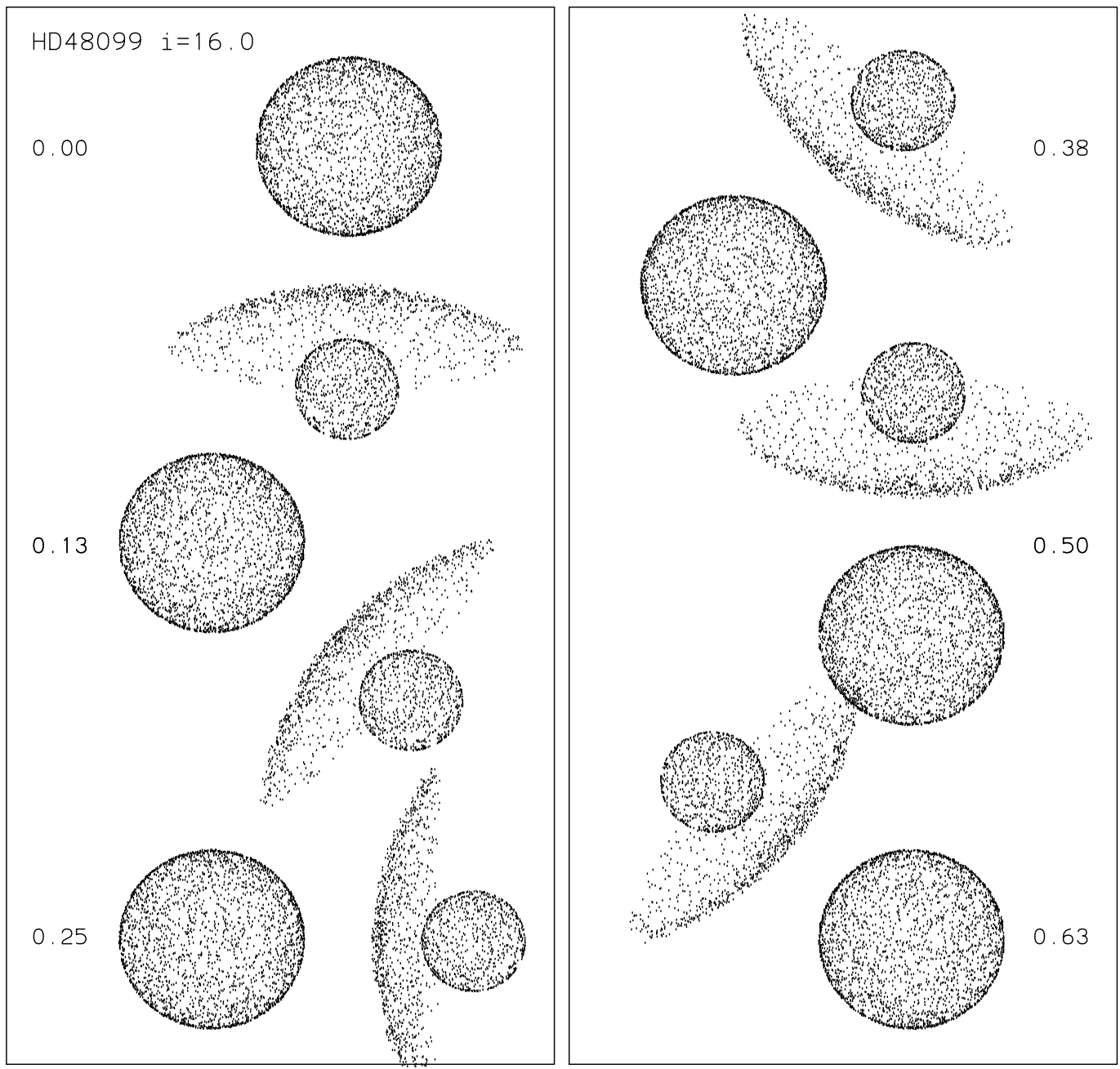}
\caption{System HD 48099 with the light scattering cloud placed at the shock produced by colliding winds as seen to an observer at different phases of the orbital period. The inclination of the orbit is $16\degr$. Photospheric scattering was assumed to be negligible for fitting the colliding winds model. }
\label{fig:windcol}%
\end{figure*}      

Light curve (count rate as function of the orbital phase) of HD 48099 is presented in Fig.~\ref{lc48}. Some variability is clearly seen, despite of the low signal-to-noise ratio and sparseness of data. There is evidence that the flux drops near the conjunction phases (0.0, 0.5) and increases at the elongations (0.25, 0.75). However, in order to confirm this, more observations in the X-rays are needed.  

A detailed spectral analysis is not possible due to the low count
statistics. However, averaging over all four observations allowed us
to determine some spectral parameters. The obtained spectrum was
grouped to have at least one count per bin$^{}$ using {\sc grppha} tool
from the {\sc FTOOLS} package and was fitted in the {\sc XSPEC} package
\citep{1996ASPC..101...17A} using Cash statistic
\citep{1979ApJ...228..939C}. X-rays from massive O+O star binaries are
emitted by the shock-heated plasma in the wind–wind collision region
\citep{pittard10}.  Therefore, this emission is
expected to be fitted with the thermal plasma models (e.g. {\sc APEC}
 model in  {\sc XSPEC}).

One-component {\sc APEC} thermal plasma model approximates the
resultant spectrum relatively well and gives the plasma temperature
$kT=0.51^{+0.12}_{-0.08}$~keV assuming solar metal abundance
\citep{1989GeCoA..53..197A}. However, inclusion of a second component
to the fitting model improves approximation quality
significantly. Namely, a C-statistic value of 41 (for 34 degrees of
freedom) for the the one-component model is reduced to 33 (for 32
degrees of freedom) for the two-component model.

In the two-component model the best-fit temperatures of two components are
$0.41\pm0.08$ and 1.7$^{+4.2}_{-0.4}$ keV. The averaged {\it Swift}/XRT spectrum of HD 48099 is shown in Figure~\ref{x-rays_sp} along with the best fit model (solid line). The soft and hard components of the model are represented by long-dashed and short-dashed lines, respectively. The model flux of
$2.5\times10^{-4}$ ph cm$^{-2}$ s$^{-1}$ ($3.1\times10^{-13}$ erg
cm$^{-2}$ s$^{-1}$) in the 0.3--5.0 keV energy range corresponds
to the luminosity $1.2\times10^{32}$ ergs s$^{-1}$ (assuming distance
of 1.8 kpc). Both plasma temperature and luminosity obtained from our
spectral analysis are typical for the OV+OV binaries as follows from
the observations \citep{2011ApJS..194....7N} and theory
\citep{pittard10}.
               
\section{Polarization model}   

In many cases, straightforward analytical treatment can be used to derive orbital parameters from the coefficients of the Fourier fit to the variations of Stokes parameters \citep{bme78,drissen86}. 
However, this method is based on certain assumptions and has some limitations. It assumes point-like stellar sources, small optical thickness of the envelope and neglects absorption processes. There is also a bias in the derived orbit inclination $i$ \citep{aspin81,simmons82,wolinski94}. As a result of the presence of noise in polarization data,  the orbit inclinations derived from the Fourier fit are systematically biased towards higher values. The magnitude of bias depends not only on the level of noise, but also on the true inclination: the lower the true value of $i$, the larger is the bias \citep[see][]{simmons82}. The effect of the stochastic noise (i.e. due to a non-periodic component in variable polarization) on the inclination derived from the BME model has been evaluated by  \citet{manset2000}. It has been shown that even in the case of high-precision measurements, this noise  alone effectively prohibits recovering inclination from the BME model for the values of $i \la 45\degr$. 

Numerical modelling has the obvious advantage that it uses a direct fit to the data and allows explicit inclusion in the model of various scattering components: for example, stream, disks, and shells. We have modelled polarization in HD 48099 with a scattering code that is based on the approach outlined in \citet{piirola80},  \citet{piirola05}, and \citet{piirola06}. Single scattering of photons by an optically thin medium, and spherical
illuminating stars with linear limb darkening coefficient, $u_r$ = 0.2, (see e.g. \citet{heyrovsky07}, Figure 2), are assumed. Changing $u_r$ in the range 0.15 - 0.35 had no significant effect on the modelled polarization 
results.          
                  
For HD 48099 we considered three possible scattering scenarios: 
\begin{enumerate}
\item reflection of light from the stellar surfaces, 
\item extended, geometrically thin scattering cloud between the stars produced by the colliding stellar winds, and 
\item a narrow accretion stream between the components. 
\end{enumerate}

We made model fits for each of these three models separately, assuming one dominant scattering component. With the available data, and without eclipses, it is impossible to simultaneously fit more than one component. There would be too many free parameters and the fits would not converge in a meaningful way.

In our models we have fixed the stellar and orbital parameters to those given in Section 1.1. When fitting our model to different inclinations we changed the component separation (orbital axis $a$) to keep our model 
consistent with the known mass function from spectroscopy. The results of our modelling for each scattering scenario are combined in Table 4. This table gives the weighting parameter, $W$ (which is proportional to the total number of scattering electrons), the parameters of the interstellar polarization 
determined by the model, ($q_{0}$, $u_{0}$), orbital inclination, $i$, the longitude of ascending node, $\Omega$, and the value of the RMS scatter, RMSD, of observed points from the fitted model curves. These results are presented separately for each passband.

The error estimates of the parameters given in Table 4 are provided by the least squares fitting procedure described in detail in Appendix of \citet{piirola05}. The best fitting inclination was found by running the
model code for a range of inclinations, and then choosing the inclination where the RMS scatter reaches the minimum. We did this for a large range of inclinations ($10\degr-24\degr$) and show in Table 4 only the results close to the best fitting inclination. 

To estimate the errors for the orbital inclination we added Gaussian noise to the polarization curves, calculated by the wind collision cloud model for a given inclination, and then refitted the model to the `noise-added' curves. From 20 fittings with different sets of noise patterns, and the same RMS and number of observations as our real data we got standard deviation of the obtained inclinations, $\sigma = 1.6\degr$  in the B and V bands, and $1.8\degr$ in the R band. These are well in accordance with the range of best fitting inclinations we got in the BVR passbands (Table 4). 

The inclinations obtained for the wind collision cloud in the BVR passbands are $17.0\degr$, $18.0\degr$, and $16.5\degr$, respectively. The independent datasets in the BVR passbands provide estimates of the `internal consistency' of each model. Since some of the RMS deviations (Table 4) is due to imperfections of the model itself, we do not assume that the estimated error of the mean inclination from the BVR bands is smaller than for each of the values obtained in the BVR passbands. Accordingly, we make an estimate $i = 17\degr \pm 2\degr$ from our colliding wind cloud model fits. The scattering shell and accretion stream models give fits that are almost as good (RMS scatter values in Table 4) as the wind collision cloud. Below we discuss each model separately. 

\begin{table*}
\caption{Best fits to the data obtained for three scattering
models.} \label{Fit_results} \centering
\begin{tabular}{c c c c c c c}
\hline \hline
   & $W$ & $q_0$ (\%) & $u_0$ (\%) & $i$(deg) & $\Omega$(deg) & RMSD (\%) \\
\hline
\multicolumn{7}{c}{Reflecting shells: A - primary star, B - secondary star} \\
B       & A: 9.77 $\pm$ 2.90 & 0.1647 $\pm$ 0.0014 & 0.0753 $\pm$ 0.0014 & 15 & 83.3 $\pm$ 0.8 & 0.00677 \\
        & B: 7.49 $\pm$ 0.99 & & & & & \\
V       & A: 6.75 $\pm$ 2.50 & 0.1744 $\pm$ 0.0012 & 0.0632 $\pm$ 0.0012 & 15 & 82.7 $\pm$ 0.8 & 0.00582 \\
        & B: 6.76 $\pm$ 0.85 & & & & & \\
R       & A: 6.28 $\pm$ 1.90 & 0.1632 $\pm$ 0.0012 & 0.0533 $\pm$ 0.0012 & 16 & 81.5 $\pm$ 1.0 & 0.00535 \\
        & B: 4.45 $\pm$ 0.64 & & & & & \\
\hline
\multicolumn{7}{c}{ Scattering cloud due to wind collision  } \\
            & 0.282 $\pm$ 0.009 & 0.1628 $\pm$ 0.0014 & 0.0760 $\pm$ 0.0014 & 15 & 83.4 $\pm$ 0.8 & 0.00677 \\
            & 0.292 $\pm$ 0.009 & 0.1629 $\pm$ 0.0014 & 0.0759 $\pm$ 0.0014 & 16 & 83.4 $\pm$ 0.8 & 0.00672 \\
            & 0.298 $\pm$ 0.009 & 0.1632 $\pm$ 0.0014 & 0.0758 $\pm$ 0.0014 & 17 & 83.4 $\pm$ 0.8 & 0.00671 \\
B           & 0.307 $\pm$ 0.009 & 0.1635 $\pm$ 0.0014 & 0.0757 $\pm$ 0.0014 & 18 & 83.4 $\pm$ 0.8 & 0.00673 \\
            & 0.319 $\pm$ 0.009 & 0.1638 $\pm$ 0.0014 & 0.0756 $\pm$ 0.0014 & 19 & 83.4 $\pm$ 0.8 & 0.00678 \\
            & 0.334 $\pm$ 0.010 & 0.1641 $\pm$ 0.0014 & 0.0754 $\pm$ 0.0014 & 20 & 83.4 $\pm$ 0.8 & 0.00687 \\
\hline
            & 0.237 $\pm$ 0.007 & 0.1731 $\pm$ 0.0012 & 0.0638 $\pm$ 0.0012 & 15 & 82.8 $\pm$ 0.9 & 0.00588 \\
            & 0.245 $\pm$ 0.007 & 0.1732 $\pm$ 0.0012 & 0.0637 $\pm$ 0.0012 & 16 & 82.8 $\pm$ 0.9 & 0.00584 \\
            & 0.250 $\pm$ 0.008 & 0.1735 $\pm$ 0.0012 & 0.0636 $\pm$ 0.0012 & 17 & 82.8 $\pm$ 0.8 & 0.00580 \\
V           & 0.257 $\pm$ 0.008 & 0.1737 $\pm$ 0.0012 & 0.0635 $\pm$ 0.0012 & 18 & 82.8 $\pm$ 0.8 & 0.00579 \\
            & 0.267 $\pm$ 0.008 & 0.1740 $\pm$ 0.0012 & 0.0634 $\pm$ 0.0012 & 19 & 82.8 $\pm$ 0.8 & 0.00580 \\
            & 0.280 $\pm$ 0.008 & 0.1742 $\pm$ 0.0012 & 0.0633 $\pm$ 0.0012 & 20 & 82.8 $\pm$ 0.8 & 0.00585 \\
\hline
            & 0.191 $\pm$ 0.007 & 0.1616 $\pm$ 0.0012 & 0.0539 $\pm$ 0.0012 & 15 & 81.6 $\pm$ 0.9 & 0.00532 \\
            & 0.198 $\pm$ 0.007 & 0.1617 $\pm$ 0.0012 & 0.0538 $\pm$ 0.0012 & 16 & 81.6 $\pm$ 0.9 & 0.00530 \\
            & 0.202 $\pm$ 0.007 & 0.1619 $\pm$ 0.0012 & 0.0538 $\pm$ 0.0012 & 17 & 81.6 $\pm$ 0.9 & 0.00530 \\
R           & 0.208 $\pm$ 0.007 & 0.1621 $\pm$ 0.0012 & 0.0537 $\pm$ 0.0012 & 18 & 81.6 $\pm$ 0.9 & 0.00533 \\
            & 0.216 $\pm$ 0.007 & 0.1623 $\pm$ 0.0012 & 0.0536 $\pm$ 0.0012 & 19 & 81.6 $\pm$ 0.9 & 0.00537 \\
            & 0.227 $\pm$ 0.008 & 0.1625 $\pm$ 0.0012 & 0.0535 $\pm$ 0.0012 & 20 & 81.6 $\pm$ 0.9 & 0.00544 \\
\hline
\multicolumn{7}{c}{Accretion stream}  \\
B       & 1.328 $\pm$ 0.040 & 0.1631 $\pm$ 0.0013 & 0.0765 $\pm$ 0.0013 & 16 & 90.0 $\pm$ 0.8 & 0.00690 \\
V       & 1.225 $\pm$ 0.038 & 0.1731 $\pm$ 0.0011 & 0.0642 $\pm$ 0.0011 & 15 & 90.0 $\pm$ 0.9 & 0.00604 \\
R       & 0.828 $\pm$ 0.028 & 0.1620 $\pm$ 0.0010 & 0.0542 $\pm$ 0.0010 & 17 & 87.7 $\pm$ 1.0 & 0.00536 \\
\hline
\end{tabular}

\end{table*}

\begin{figure} 
\centering
\includegraphics[width=9cm]{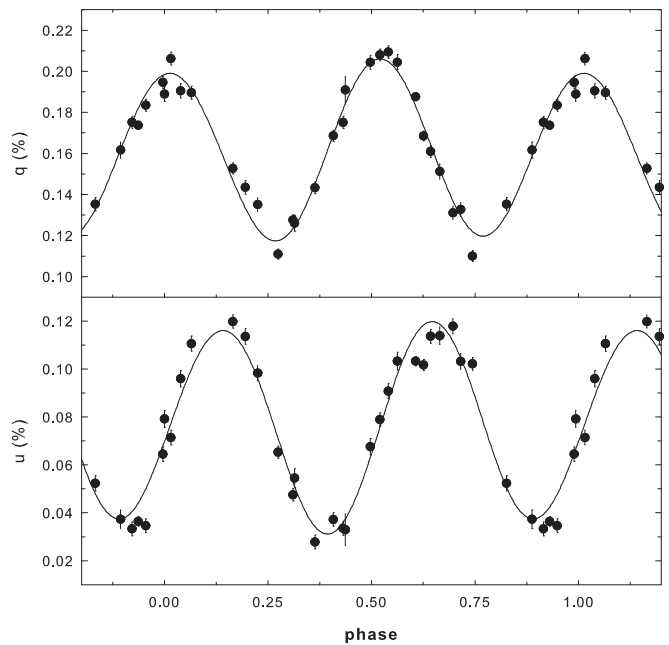}
\caption{Best model fit to the polarization data in the B-band for the colliding winds scenario.}
\label{cloud_fit}%
\end{figure}    

\subsection{Reflection from photospheres}
The binary HD 48099 is a close system and the surface of each star subtends a significant solid angle when viewed from the photosphere of its companion. Because of the high temperature, both stars have highly ionized photospheres with plenty of electron scatterers. In our model, the reflecting photospheres are approximated by thin scattering shells around the stars. The underlying photospheres are considered to be strongly absorbing. The shell layers act as scattering surfaces producing phase-dependent polarization. This approximation, while being reasonable, is no substitute for accurate and rigorous treatment of light scattering processes in stellar photospheres. Such an approach would require substantial computational efforts and is beyond the scope of this paper. However, our model reproduces the basic geometry of light scattering due to reflection effect and can be used to derive orbital parameters.
     
There are indications that the primary star in HD 48099 is a fast rotator  \citep{mahy10}. Rapidly rotating stars could be flattened. Polarization data can be used to probe this effect, because of the change in the light scattering geometry (in comparison with a spherical star) for all arbitrary inclinations, except for the extreme value of $i = 0\degr$. To investigate this possibility, we considered models with a non-spherical shell around the primary component. Our current model allows  only spherical illuminating stars. However, an optically thin ellipsoidal scattering envelope is a meaningful approximation to search for possible evidence of a (slightly) flattened structure around the stars. As we have found, the models with a shell whose equatorial radius is larger than the polar one by $5\% - 10\%$ reproduce the observed polarization variability better. The improvement is marginal, corresponding to $\sim2 \sigma$ confidence level in the least squares fitting solution, but it is persistent in all passbands for the wide range of tried inclination angles. Thus, variability of polarization in HD 48099 also favours the non-spherical shape of the primary star.
    
The best model fit to the observed variations of Stokes parameters in the V-band can be seen in Figure~\ref{shells_fit}. As we can see, the quality of fit is rather good. Thus, we can conclude that reflection mechanism seems to be capable of explaining observed variability of polarization in HD 48099. The values of the model parameters of the best fits for the B, V and R passbands are given in Table ~\ref{Fit_results}.    

\begin{figure*} 
 \centering
 \includegraphics[width=18cm]{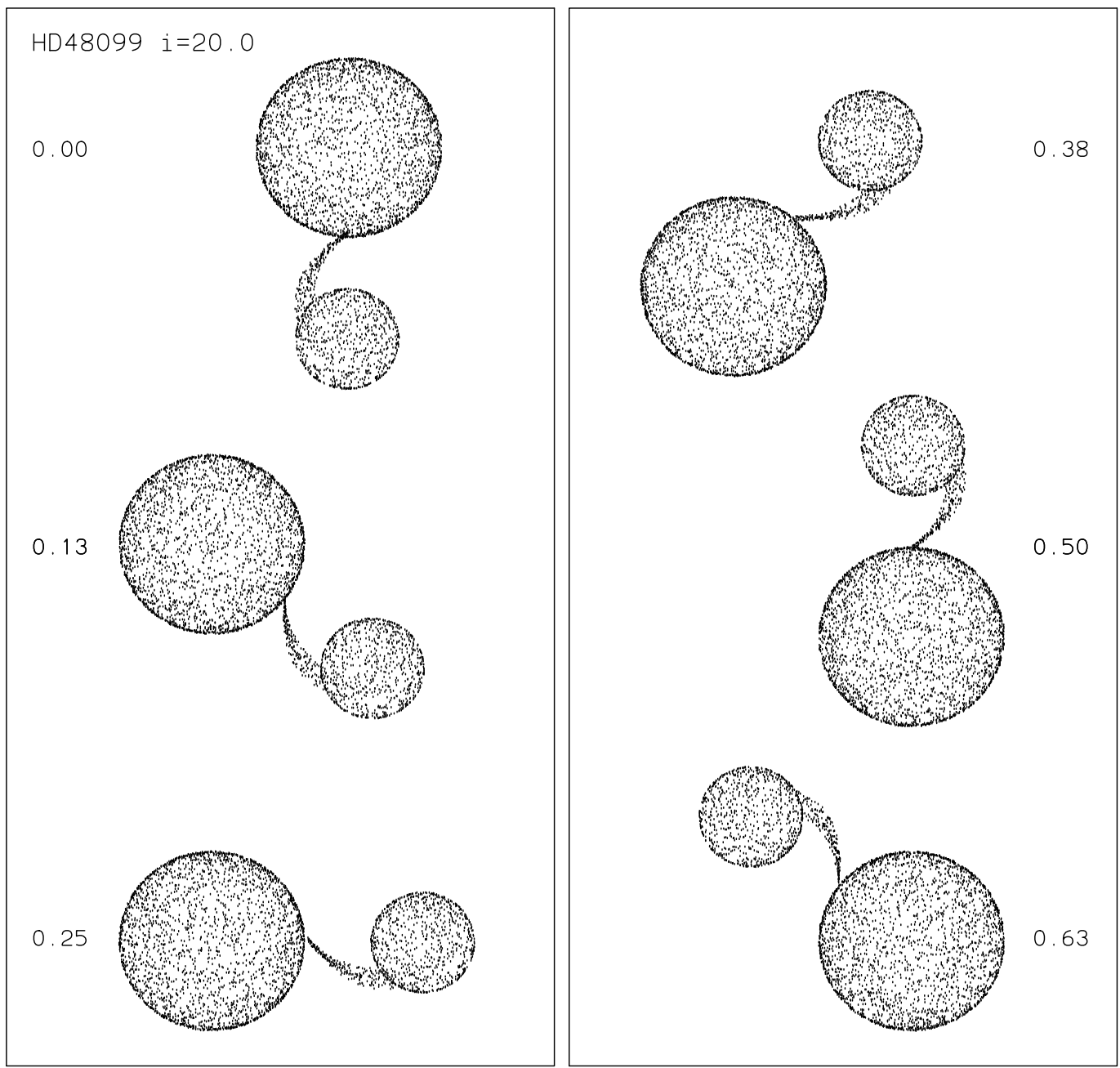}       
 \caption{System HD 48099 at inclination of the orbit  of $20\degr$ with the accretion stream between the component as seen by the observer at different phases of the orbital period. The stream is confined into the orbital plane. Photospheric scattering was assumed to be negligible for fitting the stream model. }
  \label{fig:stream}     
\end{figure*}   

\subsection{Colliding stellar winds}
In a close binary system consisting of hot luminous components the individual stellar winds may collide producing a shock area between the stars. The models of radiatively driven colliding winds in O+O binaries demonstrate that the nature of the collision region depends on the parameters of the binary system and properties of the components \citep{pittard09}.

The existence of stellar wind in HD 48099 has been revealed from the presence of P Cyg lines of N IV, N V and C IV in the UV spectra obtained with the $IUE$ \citep[see][]{howarth89}. However, there is a significant difference in the estimates of the mass loss rate by the stellar wind made by  \citet{howarth89}: $\dot M = 2 \times 10^{-7}M_{\odot}{\rm yr}^{-1}$ and \citet{mahy10}: $\dot M = 2.5 \times 10^{-8}M_{\odot}{\rm yr}^{-1}$. Both research groups apparently have used the same set of $IUE$ data to fit observed P Cyg lines, yet obtained results differ by an order of magnitude.      

The heated plasma in the wind collision area of the binary star can produce enhanced X-ray emission that can display orbital variability \citep{pittard10}. The X-ray data (see Section \ref{sec:xrays}) provide support for the wind collision region in HD 48099. Theoretical modelling also predicts the existence of such a phenomenon. HD 48099 looks very similar to the O6V+O8V system `cwb3', the hydrodynamical model which was explored by \citet{pittard09}, apart from the shorter distance between the components (see Table 2 and Figure 1(c) in \citealt{pittard09}). In our scattering model we have approximated the wind collision region in the form of a thin extended cloud located between the components. The surface of the cloud was set slightly bent to account for downstream curvature due to Coriolis forces. We also placed the cloud closer to the secondary star whose wind is weaker. We must note that our model does not allow us to put strict constraints on the distance and the extension (width) of the cloud. Similar RMS scatter values can be obtained for different combinations of the  distances to the cloud  and its width. We must also note that by increasing the inclination the distance between the component is reduced. This most likely will affect the shape and the position of the shock and our approximation may become inappropriate.  

An illustration of the scattering cloud model is shown in Figure~\ref{fig:windcol}. It is a true picture of the model, but with the number of scattering points  reduced to avoid `overexposure' of the image. The colliding wind shock model is approximated by a spherical shell with the inner and outer radii, $R_{s}$ and $R_{o}$. The center of curvature is shifted by the amount, $C_{x}$, from the center of the primary star along the line joining the stellar centers towards the secondary star to locate the cloud in the desired position. The cloud center is defined by the spherical coordinates ($\theta_c, \phi_c$). The longitude $\theta$ ($0-360\degr$) is measured in the orbital plane and has the origin in the direction of the center of the secondary seen from the center of the primary star. The latitude $\phi$ ($-90\degr - +90\degr$) is positive towards the observer. In the tangential directions the cloud was constructed by distributing the points evenly on the surface of a spherical cap of full opening angle, $\alpha$, as seen from the center of curvature. In the radial direction the cloud is filled by randomizing the distance, $r_i$, of each scattering point from the center of curvature by $r_i = R_s + x_i (R_o - R_s)$ where $x_i$ is a random real number between 0.0 and 1.0. In units of the primary star radius, $R_A$, we have applied $R_s$ = 3.3, $R_o$ =3.4, and $C_x$ = 5.2 for the model shown in Figure~\ref{fig:windcol} ($i=16\degr, \alpha = 80\degr$), to approximate the published wind collision model `cwb3', in Figure 1c from \citet{pittard09}.

Analytic expressions have been derived for the shape of the colliding-wind bow shock \citep{canto96}, and for the asymptotic opening angle \citep{Gayley09}. The approximate opening angle $\sim60\degr$ of the outflow in our model would correspond to wind momentum ratio $\sim 0.15$ for radiative shocks, \citep[see Figure 2,][]{Gayley09}. For smaller wind momentum ratios the inferred shock cone opening angle would decrease and the shock move closer to the weaker wind star \citep{canto96}.

A detailed modelling of the wind shock region is beyond the scope of the present paper, but our model of the curved cloud, extended into a large region away from the orbital plane, approximates the scattering region in a meaningful way for searching effects from such geometry on phase-locked polarization over the binary period. Our model represents the region where the dominant polarized flux arises. With increasing distance from the stars the polarized flux decreases rapidly due to decreasing illumination and electron density.

An example of the colliding wind model fit to the polarization data can be seen in Figure~\ref{cloud_fit}.
The model with the scattering of light from the shock region reproduces the observed variability of polarization in HD 48099 quite well. The model parameters  which produce the best fit are given in Table~\ref{Fit_results}. We note that the best fit was obtained by shifting the cloud center to $\phi_{c} = 2\degr$ which corresponds to a displacement by $0.1R_{A}$ above the orbital plane (towards the observer). This could be understood as being due to optical thickness effects in the cloud: the photons scattered from regions above the orbital plane can reach the observer more easily than below the plane, and the `effective' center of the scattering cloud is shifted towards the observer. However, the low optical thickness of the cloud ($\tau < 0.01$, Sect. 5) does not support this interpretation, leaving the possibility that a complex cloud geometry may be responsible for the displacement.
\subsection{Accretion stream}
As mentioned by \citet{mahy10}, for the sizes and masses of the components derived for HD 48099, the primary could fill its Roche lobe at the inclination of about $20\degr$. In this case the accretion stream from the primary to the secondary may appear due to the Roche lobe overflow. Although there is no evidence for the presence of ongoing mass exchange in HD 48099, we have explored this scattering mechanism in our model fits. 

\begin{figure} 
 \centering
 \includegraphics[width=9cm]{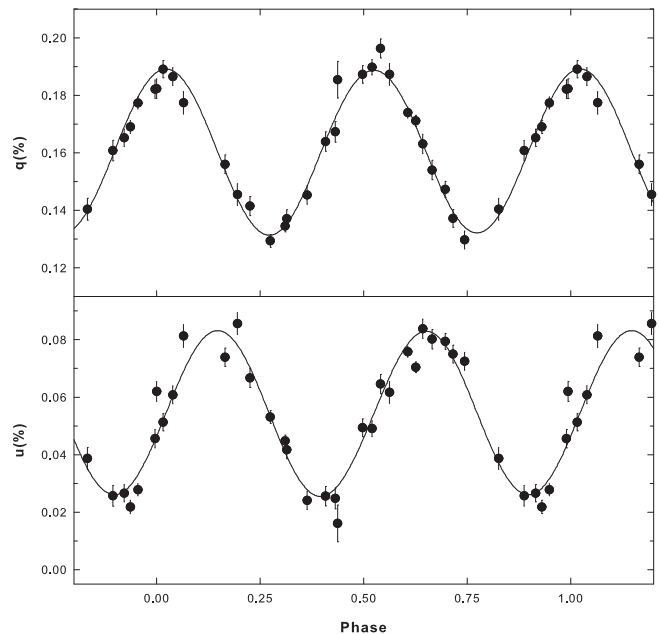}
 \caption{Best model fit to the polarization data in the R-band for the accretion stream scenario.}
 \label{stream_fit}%
\end{figure}

 An illustration showing the system HD 48099 with the narrow accretion stream directed from primary to secondary is shown in Figure~\ref{fig:stream}. The starting point is on the primary surface at the line connecting the stars. The stream is deflected from the line connecting the stellar centres by the Coriolis force towards the trailing side of the star B. The end (impact) point is at the distance $0.8 R_B$ from the line connecting the stars. A constant number of scatterers per unit length of the stream is assumed. We added some curvature (Figure~\ref{fig:stream}) to the trajectories by fitting a second order polynomial curve through three points, the central one deviating from a straight line between the start and end point by 0.2 of the distance between the start and end point. We also added spreading of the stream, $dw = 0.2(y_i - 0.5)s$, where $y_i$ is a random real number between 0.0 and 1.0, and $s$ is the distance of the scattering point from the starting point of the stream. The stream is confined into the orbital plane. We have tried several models with various degrees of deflection and widths of the stream. The shape of the stream has very little effect on polarization in a non-eclipsing system, and cannot be determined from our observations. The model fit to the polarization data in the V-band is shown in Figure~\ref{stream_fit}. Model parameters of the best fits are given in Table~\ref{Fit_results}. As with the other scattering scenarios, a stream model represents observational data quite well.

\section{Comparison of scattering models}

The results of our modelling with three different scenarios allow us to draw some important conclusions: \begin{enumerate}
 \item The colliding wind and scattering shells models explain the observed polarization variability equally well. However, the `shells' model has one more free parameter: the number of scatterers (weighting factor $W$) is determined both for the shell around the primary (A) and secondary star (B). Because the available X-ray data support the existence of colliding stellar winds, this can be, in principal, considered as a decisive factor. However, our models are just approximations of more complex scattering mechanisms: each of them requires more detailed and rigorous consideration. The main problem in interpretation of variable polarization in HD 48099 is the low inclination of the binary orbit and absence of eclipses. In binary systems seen at high inclinations, obscuration of the scattering material occurring at certain orbital phases would help to locate the main source of the scattered radiation. 
             
 \item The stream model fit is slightly worse, particularly in the B and V bands (Table 4) in comparison with the reflection and the scattering cloud models. Because of this fact, and also because there is no observational evidence for the ongoing mass outflow in HD 48099, scattering on the accretion stream seems unlikely to be responsible for the observed polarization variability. However, it is interesting to note that the X-ray data (Figure 3) show a maximum in the phase interval where the stream impact region is best visible (Figure~\ref{fig:stream}).
      
 \item The orbital parameter $\Omega$, derived for the reflection and scattering cloud models are similar. For all passbands, the best fits are achieved for orbit orientation $\Omega = 81\degr - 83\degr$. In the case of the stream model, there is a systematic difference in derived value of $\Omega$, which is higher ($\sim90\degr$). This is due to the fact that the orbit orientation obtained from the model depends on the longitude of the scattering cloud (or stream). The orientation of the scattering material projected on the sky is well defined by the polarization curves. If the location of the material relative to the binary components is changed, $\Omega$ will also immediately change. Even if we cannot make a clear decision on the scattering mechanism working in HD 48099, we can conclude that orbit orientations obtained from polarization data in different passbands are consistent for the shells and wind collision cloud models and depend somewhat on the model choice for the stream model.   

 \item The colliding wind cloud model gives orbital inclinations with good consistency in different passbands: $i_{B} = 17\degr$, $i_{V} = 18\degr$, and $i_{R} = 16.5\degr$. This is in accordance with our Gaussian noise error estimates for inclination (Sect. 4), $\sigma_{B} = 1.6\degr$, $\sigma_{V} = 1.6\degr$, and $\sigma_{R} = 1.8\degr$, in the BVR passbands, respectively. The inclination determinations are obviously model dependent, and we cannot put strict limits on their uncertainties. However, we have tried a plane layer cloud between the stars, and a cloud with opposite curvature, and got similar inclinations, within the quoted ($\pm2\degr$) errors. The phase-locked polarization is not strongly dependent on the exact geometric structure, and does not provide means for a detailed comparison of published shock front models, \citep[see e.g.][] {Gayley09,Usov92}. The essential feature is that we have a broad distribution of scatterers away from the orbital plane. Our model is not sensitive to instabilities of the shock front geometry. Polarization variations due to changes in the scattering cloud would be seen as deviating points in our model fit curves, and their effects thereby included in the estimates of uncertainty from the least squares model fit. The Gaussian noise error analysis for the shells and the stream models shows that they do not provide reasonable constraints for inclination in the  range $i = 10\degr-25\degr$. For the shells this is because the relative electron densities of the two shells adjust for the inclination changes and the RMS scatter of the best fit depends only weakly on inclination. For the stream, the inclination dependence of the model fits is weak because the stream is confined in the orbital plane. Then even very good data are not enough to provide good constraints below $i \sim 30\degr$. This is in accordance with the earlier results by \citet{wolinski94} and \citet{manset2000}. In our colliding wind model the scattering matter is distributed in a wide region away from the orbital plane. This also puts good constraints on inclination at low inclinations ($10\degr < i < 25\degr$).
   
\end{enumerate}         

The weighting factor $W$ illustrates how strongly the geometry of the scattering material affects the efficiency in producing polarization. In the case of shells, a large fraction of scattering particles is obscured by the stellar body. For the stream, the scattering angles for the radiation from the secondary are such that the resulting polarization tends to cancel that of the scattered flux produced by the primary. The wind collision cloud is well visible throughout the orbital cycle and the polarized fluxes co-add in the regions where the illumination by both component stars is strongest. Therefore, a much smaller number of scatterers is sufficient to produce the observed polarization. Assuming a radius of $11.6 R_{\odot}$ for the primary star, the weighting factor $W$ corresponds to $m \simeq 0.7 \times 10^{-11}M_{\odot}$ of fully ionized hydrogen for the wind collision cloud (see Equations 2 and 3 in \citealt{piirola05}). This is $\sim 3\times10^{-4}$ of the annual mass loss estimated by Mahy et al. (2010) and $\sim 3\times10^{-5}$ of that estimated by Howarth \& Prinja (1989). The total number of scattering electrons given by our model, $N_e = 0.8\times10^{46}$, and the adopted dimensions of the scattering cloud (Sect. 4.2), give an average electron density in the cloud, $n_e = 1.2\times10^{10}cm^{-3}$. From the well known dependence, $\tau = \sigma_o n_e s$, where $\sigma_o$ is the Thomson scattering cross section, and $s$ is the path length, we can estimate the optical thickness $\tau$. The cloud is optically thin, $\tau < 0.01$, even when viewed 
sideways (Figure 6, $s \sim 2R_{A}$).  

\section{Conclusions}

High precision polarization measurements of the early-type binary HD 48099 have revealed variable phase-locked polarization with the amplitude at the level of $\leq 0.1\%$. From the model fits to the data we have found that light scattering on the cloud produced by colliding stellar wind and/or reflection of light from the stellar photospheres  are the most probable sources of the polarized radiation in the system. Although the available X-ray data give evidence for the existence of the wind collision region in the system, we conclude from our analysis that both mechanisms could explain the polarization variability. In order to make a definite decision, more rigorous treatment of the light scattering and wind collision processes is needed. 

Good constraints on the orbit inclination $i = 17\degr \pm 2\degr$ and orientation $\Omega = 82\degr \pm 1\degr$ are put by the colliding stellar wind model. However, the models with circumstellar envelopes or the accretion stream do not provide good constraints on the low inclination, better than suggested by the absence of eclipses ($i \le 27\degr$). The orbital parameter $\Omega$ is not affected significantly by the choice of the scattering model. Our results clearly demonstrate that high-precision polarimetry can be used as an effective tool for studying early-type binary systems.
 
The variable X-ray flux detected in HD 48099 by the  {\it Swift} observatory makes this system a promising target for future observations with the X-ray telescopes. High-precision X-ray data obtained with good orbital phase coverage can provide important information on the properties of the wind collision region in this system. These data can be also useful for better understanding of polarization mechanisms that operate in early-type binary systems.   

\begin{acknowledgements}
 This work was supported by the ERC Advanced Grant HotMol ERC-2011-AdG-291659 (www.hotmol.eu). DIPOL-2 was built in the cooperation between the University of Turku, Finland, and the Kiepenheuer Institut fuer Sonnenphysik, Germany, with the support by the Leibniz Association grant SAW-2011-KIS-7. We are grateful to the Swift team for the execution of our ToO request. JP acknowledges support from the Academy of Finland grant 268740. We are grateful to the Institute for Astronomy, University of Hawaii (IfA), for the allocation of IfA time for our observations at the T60 telescope. This research has made use of the Simbad database, operated at CDS, Strasbourg, France. 
\end{acknowledgements}


\end{document}